\documentclass[twocolumn,showpacs,preprintnumbers,amsmath,amssymb]{revtex4}
\input psfig.sty
 
\usepackage{graphicx}
\usepackage{dcolumn}
\usepackage{bm}
 
\begin{document}
 
\def\be{\begin{equation}}
\def\ee{\end{equation}}
 
 
 
\title{The $H$-theorem in $\kappa$-statistics: influence on the
molecular chaos hypothesis}

\author{R. Silva\footnote{On leave of absence from Universidade do
Estado do Rio Grande do Norte, 59610-210, Mossor\'o, RN, Brasil}} \email{rsilva@on.br,rsilva@uern.br}
\affiliation{Observat\'orio Nacional, Rua Gal. Jos\'e Cristino 77,
20921-400 Rio de Janeiro - RJ, Brasil} 


\date{\today}
 
\begin{abstract}
We rediscuss recent derivations of kinetic equations based on the
Kaniadakis' entropy concept. Our primary objective here is to derive a kinetical
version of the second law of thermodynamycs in such a $\kappa$-framework. To this end, we
assume a slight modification of the molecular chaos hypothesis. For the $H_{\kappa}$-theorem, it is shown
that the collisional equilibrium states (null entropy source term)
are described by a $\kappa$-power law extension of the
exponential distribution and, as should be expected, all these
results reduce to the standard one in the limit $\kappa
\rightarrow 0$. 
\end{abstract}
 
\maketitle
 
\section{introduction}
 
Over the last few years, a great deal of attention has been paid
to nonextensive statistic mechanics  based on the deviations
of Boltzmann-Gibbs-Shannon entropic measure. Basically, in this
extended framework, the key point is to substitute the exponential behaviour of the entropy by a
power-law one (see, e.q. \cite{T95b}). Recently, similar motivations also led at least to two new examples, namely, Abe \cite{abe97,abe2004} and Kaniadakis entropies \cite{k1,k2}. In this latter works, by using $\kappa$-exponential and $\kappa$-logarithm functions (see Eqs. (\ref{expkappa}) and (\ref{lnkappa})) and the kinetic interaction principle (KIP), Kaniadakis proposed a statistical framework based on $\kappa$-entropy \cite{k1,k2}
\begin{equation}\label{e1}
S_{\kappa}(f)=-\int d^{3}v f [a_{\kappa}f^{\kappa}+a_{-\kappa}f^{-\kappa}+b_{\kappa}]
\end{equation}
where $a_{\kappa}$ and $b_{\kappa}$ are coefficients so that in limit $\kappa\rightarrow 0$, Eq. (\ref{e1}) reduces to the standard entropy $S_{\kappa=0}$. Expression (\ref{e1}) is also the most general one that leads to the $\kappa$-framework.

Previous works have already discussed some specific choices for the constants $a_\kappa$ and $b_\kappa$. For instance, for the pair [$a_\kappa = 1/2\kappa$, $b_\kappa=0$], it is possible to write the Kaniadakis entropy as \cite{k1,abe2004}
\begin{equation}
S_\kappa = - \int d^3 v f\ln_\kappa f =  - \langle{\ln_\kappa (f)\rangle},
\end{equation}
which has a perfect analogy with the standard formalism. The second and third choices are, respectively, [$a_\kappa= 1/2\kappa(1+\kappa)$, $b_\kappa=0$] \cite{k1} and [$a_\kappa= 1/2\kappa(1+\kappa)$, $b_\kappa=-a_\kappa-a_{-\kappa}$] \cite{abe2004} while the fourth one is given by [$a_\kappa= Z^\kappa/2\kappa(1+\kappa)$, $b_\kappa=0$] \cite{k1}. In particular, in
this latter choice, the $\kappa$-entropy is given by
\cite{k1,kaniad2001}
\begin{equation} \label{tsaent}
S_{\kappa} = - \int{d^{3}v
            \left(\frac{z^{\kappa}}{2\kappa(1+\kappa)}f^{1+\kappa}
                 -\frac{z^{-\kappa}}{2\kappa(1-\kappa)}f^{1-\kappa}\right)}.
 \end{equation}
The $\kappa$-statistic is defined by the
$\kappa$-deformed functions, given by
\begin{equation}\label{expkappa}
\exp_{\kappa}(f)= (\sqrt{1+{\kappa}^2f^2} + {\kappa}f)^{1/{\kappa}},
\end{equation}
\begin{equation}\label{lnkappa}
\ln_{\kappa}(f)= {f^{\kappa}-f^{-\kappa}\over 2\kappa},
\end{equation}
and
\begin{equation}\label{Inv}
\exp_\kappa(\ln_\kappa f) = \ln_\kappa(\exp_\kappa(f))= f.
\end{equation}
As one may check, the above functions reduce to the standard exponential and logarithm when
$\kappa\rightarrow 0$. In particular, this $\kappa$-framework
leads to a class of one parameter deformed structures with
interesting mathematical properties \cite{kaniad2001}. Recently, a
connection with the generalized Smoluchowski equation was
investigated \cite{chava04}, and a fundamental test, i.e., the so-called
Lesche stability was also checked in the $\kappa$-framework
\cite{kaniadakis04}. More recently, it was shown that it is possible to obtain a consistent form
for the entropy (linked with a two-parameter deformations of logarithm
function), which generalizes the Tsallis, Abe and Kaniadakis logarithm behaviours \cite{kania05}. In the experimental viewpoint, there exist some evidence
related with the $\kappa$-statistic, namely, cosmic rays flux,
rain events in meteorology \cite{kaniad2001}, quark-gluon plasma \cite{miller03}, kinetic models describing a gas of interacting atoms and photons \cite{rossani04}, fracture
propagation phenomena \cite{cravero04}, and income distribution \cite{drag03}, as well as construct financial models \cite{bolduc05}. In the theoretical front, some studies on the canonical quantization of a classical system has also been investigated \cite{scarfone05}.
 
In this letter, we aim at rediscussing the $H$-theorem in the context of Kaniadakis entropy framework, Eq. (\ref{tsaent}). However, instead of using the KIP introduced in Ref. \cite{kaniad2001}, we propose a different route to the $\kappa$-statistic which follows similar arguments of Ref. \cite{lima01}. In reality, the main result is to obtain the equilibrium velocity $\kappa$-distribution of a slight modification of the kinetic Boltzmann $H$-theorem, where the central idea follows by modifing the molecular chaos hypothesis and generalization of the local entropy formula in accordance with a $\kappa$-statistic. 

\section{The molecular chaos hypothesis}
 
As is widely known, the Boltzmann's kinetic theory (BKT) relies on two statistical ingredients, namely
\cite{Sommerf,Colin}:
\begin{itemize}
 
\item The local entropy which is expressed by Boltzmann's
logarithmic measure (Boltzmann's constant is an unity)
 
\be \label{Boltzmann1} H[f] \, = \, - \, \int \, f({\bf r}, {\bf v}, t) \, \ln f({\bf r}, {\bf v}, t) \, d^3v. \ee
 
\item The hypothesis of molecular chaos (``Stosszahlansatz"),
i.e., the two point correlation function of the colliding particles can be factorized
 
\be \label{Boltzmann2} f({\bf r}_1, {\bf v}_1, {\bf r}_2, {\bf v}_2,
t) \, = \, f({\bf r}_1, {\bf v}_1, t) \, f({\bf r}_2, {\bf v}_2, t). \ee
\end{itemize}
There is some controversy associated with the second assumption, Eq. (\ref{Boltzmann2}). In particular, Burbury \cite{B1894}, was the first to point out that this hypothesis provides the fundamental role within the BKT. Physically, the equation (\ref{Boltzmann2}) represents that colliding molecules
are uncorrelated, i.e., the velocities and positions of pairs of
molecules are statistically independents. The irreversibility associated with the Boltzmann's equation can be traced back to this assumption. Indeed, as the molecules which has
been assumed to be uncorrelated {\it before} a collision,
become correlated {\it after} the collision, this clearly represents a time asymmetric hypothesis \cite{Z92}. In particular, this hypothesis may not always be valid
for real gases (see \cite{B83,B64} for details).

In this investigation, we introduce a consistent generalization of this
hypothesis, expression (\ref{Boltzmann2}), within the $\kappa$-statistic proposed by
Kaniadakis. We remark that equation (\ref{Boltzmann2}) implies
that the logarithm of the joint distribution $f({\bf r}_1, {\bf v}_1, {\bf r}_2, {\bf v}_2, t)$ describing coliding molecules is given by
\begin{equation}
\ln f({\bf r}_1, {\bf v}_1, {\bf r}_2, {\bf v}_2,
t) \, = \, \ln f({\bf r}_1, {\bf v}_1, t) + \ln f({\bf r}_2, {\bf v}_2, t)
\end{equation}
where each term involve only the
marginal distribution associated with one of the colliding
molecules. The new hypothesis assumed here is to
consider that  {\it a power} of the joint distribution ({\it instead
of the logarithm}) be equal to the sum of two terms, each one
depending on just one of the coliding molecules. Considering the $\kappa$-logarithm function, the condition above can be formulated in a way that extend the standard
hypothesis, Eq. (\ref{Boltzmann2}).

\section{H-theorem and $\kappa$-statistics}

Let us now introduce a spatially homogeneous gas of $N$ hard-sphere
particles of mass $m$ and diameter $s$, under the action of an
external force ${\bf F}$, and enclosed in a volume $V$. In BKT the state of a non-relativistic gas is characterized by the
one-particle distribution function $f({\bf r},{\bf v},t)$, which
is defined in such a way that $f({\bf r},{\bf v},t)d^{3}{\bf r} d^{3}{\bf v}$
gives at a time t, the number of particles in the volume element
$d^{3}{\bf r}d^{3}{\bf v}$ around the particle position ${\bf r}$ and velocity
${\bf v}$. Let us consider that distribution function is a solution
of the $\kappa$-Boltzmann equation, given by
\begin{equation}
\label{Beq} \frac{\partial f}{\partial t} +
                    {\bf v}\cdot\frac{\partial f}{\partial {\bf r}} +
                    \frac{\bf F}{m}\cdot\frac{\partial f}{\partial
{\bf v}}
                    = C_{\kappa}(f) ,
\end{equation}
where $C_{\kappa}$ defines the $\kappa$-collisional term. In expression (\ref{Beq}), the
left-hand-side (LHS) is just the total time derivative of the distribution function, thus it is reasonable to consider that the unique possible modification describing $\kappa$-statistic must be associated with the
collisional term, which is clearly a hypothesis of work. Basically, $C_{\kappa}(f)$ may be calculated through the laws of elastic collisions, where the standard assumptions are also valid \cite{Colin,Sommerf}.
 
As in the canonical $H$-theorem, our main
goal here is to show that $C_{\kappa}(f)$ leads to a nonnegative
expression for the time derivative of the $\kappa$-entropy, and does not vanish unless
the distribution function assumes the equilibrium form for a
$\kappa$-distribution which has been recently proposed \cite{kaniad2001}.
Here, we define
\begin{equation} \label{eq:2.11}
C_{\kappa}(f)={s^2\over 2}\int |{\bf V} \cdot {\bf e}| R_{\kappa}
d\omega d^{3}v_1,
\end{equation}
where $d^{3}v_1$ is an arbitrary volume element in the velocity
space, ${\bf V}$ denotes the relative velocity before collision,
${\bf V}= {\bf v_1} - {\bf v}$, ${\bf e}$ denotes an arbitrary
unit vector, $d\omega$ is an elementary solid angle such that
$s^2d\omega$ is the area of the ``collision cylinder" (for details
see Refs. \cite{Sommerf,Colin}), finally $R_{\kappa}(f,f')$
is a diference of two correlation functions (just before and after
collision). In this approach, such expression is assumed to satisfy a $\kappa$-generalized form of
molecular chaos hypothesis, which is given by
\begin{eqnarray*} \label{eq:2.12}
R_{\kappa} = \exp_{\kappa}\left(\ln_{\kappa}{{z'} f'}+
                          \ln_{\kappa}{{z'}_{1}
                          f'_{1}}\right)
\end{eqnarray*}
\begin{equation} -
            \exp_{\kappa}\left(\ln_{\kappa}{{z} f} +
                          \ln_{\kappa}{{z}_{1}f_{1}}\right)
,
\end{equation}
where primes refer to the distribution function after collision, $z,z'$ are arbitrary constant and $exp_\kappa(f)$, $\ln_\kappa(f)$, are defined by Eqs. (2) and (3). Note that
$\kappa\rightarrow 0$ the above expression reduces to $R_0=({z'}f')({z'_1}{f'}_1)-(zf)({z_1}f_1)$, where is one standard to the molecular chaos hypothesis. 
 
In the present framework, we adopt Kaniadakis formula for
local entropy 
\begin{equation} \label{eq:1}
  H_{\kappa} = - \int{d^{3}v
\left(\frac{z^{\kappa}}{2{\kappa}(1+{\kappa})}f^{1+{\kappa}}
-\frac{z^{-\kappa}}{2{\kappa}(1-{\kappa})}f^{1-{\kappa}}\right)}.
\end{equation}
Here, in order to obtain the source term, we need the partial time derivative of $H_{\kappa}$
\begin{equation} \label{eq:14}
  \frac{\partial H_{\kappa}}{\partial t} =
                  - \int{d^{3}v \ln_{\kappa}{fz}
                   \frac{\partial f}{\partial t}},
\end{equation}
and combining with the $\kappa$-Boltzmann equation (\ref{Beq}) and (\ref{eq:2.11}), one may rewrite the expression in the form of a balance equation
\begin{equation} \label{eq:2.19}
{\partial H_\kappa\over\partial t} + \nabla\cdot{\bf S}_k=G_{\kappa}({\bf r},t),
\end{equation}
where the $\kappa$-entropy flux vector related to $H_{\kappa}$ is given by
\begin{equation}\label{eq:2.18}
{\bf S}_{\kappa} = - \int{d^{3}v
{\bf v}\left(\frac{z^{\kappa}}{2\kappa(1+\kappa)}f^{1+\kappa}-\frac{z^{-\kappa}}{2\kappa(1-\kappa)}f^{1-\kappa}\right)},
\end{equation}
and the source term reads
\begin{equation} \label{eq:2.21}
G_{\kappa}={-s^2\over 2}\int |{\bf V} \cdot {\bf e}| \ln_{\kappa}fz
\;\; R_{\kappa} d\omega d^{3}v_1d^{3}v.
\end{equation}
At this point, it is convenient to rewrite $G_{\kappa}$ in a more
symmetrical form by using some elementary symmetry operations
which also take into account the inverse collisions. First we
notice that by interchanging ${\bf v}$ and ${\bf v}_1$ the value of the
integral is preserved. This happens because the magnitude of the relative velocity vector and the scattering cross section are invariants
\cite{Sommerf}. In addition, the value of $G_{\kappa}$ is not
altered if we integrate with respect to the variables ${\bf v}'$ and
${\bf v}'_1$. Actually, although changing the sign of $R_{\kappa}$ in this step
(inverse collision), the quantity ${d^3vd^3v_1}$ is also
a collisional invariant \cite{Sommerf}. As one may check, such considerations imply that the
$\kappa$-entropy source term can be written as
\begin{eqnarray*}  \label{eq: 2.26}
G_{\kappa}({\bf r},t)={s^2\over 8} \int |{\bf V} \cdot
{\bf e}| (\ln_{\kappa}{z'_{1}f'_{1}} +\ln_{\kappa}{z'f'}-
\end{eqnarray*}
\begin{eqnarray*}
-\ln_{\kappa}{z_{1}f_{1}} -\ln_{\kappa}{zf})[\exp_{\kappa}(\ln_{\kappa}{{z'} f'}+\ln_{\kappa}{{z'}_{1}f'_{1}})-
\end{eqnarray*}
\begin{eqnarray}\label{source}
-\exp_{\kappa}(\ln_{\kappa}{{z} f} +
\ln_{\kappa}{{z}_{1}f_{1}})] d\omega d^{3}v_1d^{3}v.
\end{eqnarray}
Note that the integrand in (\ref{source}) is never negative,
because the expressions
\begin{equation}
(\ln_{\kappa} z'f'+\ln_{\kappa} z'_{1} {f}'_1-\ln_{\kappa}
zf-\ln_{\kappa} z_{1}f_1)
\end{equation}
and
\begin{eqnarray} 
\exp_{\kappa}(\ln_{\kappa}z'f'+\ln_{\kappa}z'_{1}{f'}_1)
-\exp_{\kappa}(\ln_{\kappa}zf+\ln_{\kappa}z_{1}f_1)
\end{eqnarray}
always have  the same signs.
Therefore, for values of $\kappa$ on the interval  $[-1;1]$, we
obtain
\begin{equation} \label{eq:2.28}
{\partial H_{\kappa}\over\partial t} + \nabla\cdot{\bf S}_{\kappa}=G_{\kappa}\ge 0,
\end{equation}
which is the mathematical expression for the $H_{\kappa}$-theorem.
This inequality states that the $\kappa$-entropy source must be
positive or zero, thereby furnishing a kinetic derivation of the
second law of thermodynamics in the $\kappa$-ststistic.

Now, in order to make the H-theorem and the $\kappa$-statistic compatible, we need to recover the
related equilibrium distribution previously obtained by an
extremization of $\kappa$-entropy and KIP \cite{kaniad2001}. The $H_\kappa$-theorem states that $G_\kappa=0$ is a necessary 
and sufficient condition for equilibrium. Since the integrand of
(\ref{source}) cannot be negative, this occur if and only if
\begin{equation}
\ln_{\kappa} z'f'+\ln_{\kappa} z'_{1} {f}'_1=\ln_{\kappa}
zf+\ln_{\kappa} z_{1}f_1
\end{equation}
 
Therefore, the above sum of $\kappa$-logarithms remains constant
during a collision, or equivalently, it is a summational
invariant. Indeed, the unique quantities satisfying
(\ref{source}) are the particle masses, and the expressions for
momentum and energy conservation laws, which lead to the following
expression
\begin{equation} \label{eq:2.32}
\ln_{\kappa} zf = a_o+{\bf a}_1.{\bf v}+a_2{\bf v}^2,
\end{equation}
where $a_o$ and $a_2$  are constants and ${\bf a}_1$ is an
arbitrary constant vector. By introducing the barycentric
velocity, ${\bf u}$, we may rewrite (\ref{eq:2.32}) as
\begin{equation} \label{eq:2.33}
\ln_{\kappa} z f=\alpha-\gamma{({\bf v}-{\bf u})}^2 ,
\end{equation}
with a different set of constants. Thus, we
obtain a $\kappa$-distribution
\begin{equation} \label{eq:2.34a}
f={1\over z}\exp_{\kappa}[\alpha-\gamma
{({\bf v}-{\bf u})}^2],
\end{equation}
where  $\gamma$ and ${\bf u}$ may be functions of the
temperature. The expression above is the general form of the
$\kappa$-Maxwellian distribution function.

\section{Final remarks}

Summing up, we have discussed a $\kappa$-generalization of Boltzmann's kinetic
equation along the lines defined by $\kappa$-statistic. The main results follow from a slight
modification of the main {\it statistical hypothesis} underlying
Boltzmann's approach: 

\begin{enumerate}

\item The $\kappa$-statistic has been
explicitly introduced through a new functional formula for the
local entropy;
 
\item A nonfactorizable expression for the
molecular chaos hypothesis has been adopted. 

\end{enumerate}
Both ingredients are shown to be consistent with the standard laws describing the
microscopic dynamics, and reduce to the familiar Boltzmann assumptions in the limit ($\kappa=0$).

\noindent {\bf Acknowledgments:} The author is grateful to J. S. Alcaniz, I. Queiroz and J. A. S. Lima for helpful discussions. This work was supported by
Conselho Nacional de Desenvolvimento Cient\'{i}fico e
Tecnol\'{o}gico - CNPq (Brasil).

\end{document}